\documentclass[twocolumn,prl]{revtex4}
\usepackage{amssymb}
\usepackage{graphicx}

\input{tcilatex}

\begin{document}

\title{Diameter selective characterization of single-wall carbon nanotubes}
\author{F. Simon}
\author{\'{A}. Kukovecz$^{*}$}
\author{C. Kramberger}
\author{R. Pfeiffer}
\author{F. Hasi}
\author{H. Kuzmany}
\address{Institut f\"{u}r Materialphysik, Universit\"{a}t Wien, Strudlhofgasse 4,
A-1090 Wien, Austria}
\author{H. Kataura}
\address{Department of Physics, Tokyo Metropolitan Univ, Tokyo, Japan}

\begin{abstract}
A novel method is presented which allows the characterization of diameter
selective phenomena in SWCNTs. It is based on the transformation of
fullerene peapod materials into double-wall carbon nanotubes and studying
the diameter distribution of the latter. The method is demonstrated for the
diameter selective healing of nanotube defects and yield from C$_{70}$
peapod samples. Openings on small diameter nanotubes are closed first. The
yield of very small diameter inner nanotubes from C$_{70}$ peapods is
demonstrated. This challenges the theoretical models of inner nanotube
formation. An anomalous absence of mid-diameter inner tubes is observed and
explained by the suppressed amount of C$_{70}$ peapods due to the
competition of the two almost equally stable standing and lying C$_{70}$
peapod configurations.
\end{abstract}

\maketitle




Nanostructures based on carbon nanotubes \cite{IijimaNAT} have been in the
forefront of nanomaterial research in the last decade. However, there still
remains a number of open questions before one of the most promising
candidates, single-wall carbon nanotubes (SWCNTs) will have wide-spread
applications. The most important obstacle is the large number of
electronically different nanotubes produced with similar diameters and
different chiralities \cite{Dresselhaus}. To overcome this, recent efforts
have focused on separation of SWCNTs \cite{ChattopadhyayJACS}\cite{KrupkeSCI}%
\cite{RinzlerNL}\cite{StranoSCI}. The general problem remaining before the
successfull optimization of such methods is the identification of the
separated SWCNTs according to their chiral vectors. Another prerequisite for
the applicability of SWCNTs is the capability of studying the behavior of
different SWCNTs against chemical and physical treatments in a chirality
sensitive manner. Band-gap flurescence was successfully applied to assign
the chiral index to semiconducting SWCNTs \cite{WeismanSCI}. In a different
approach, assignment to chiral vectors of small diameter nanotubes in
double-wall carbon nanotubes (DWCNTs) was performed using Raman spectroscopy 
\cite{PfeifferPRL}\cite{KrambergerPRB}. DWCNTs are SWCNTs containg a
coaxial, smaller diameter CNT. The material is produced from fullerene
encapsulated SWCNT materials, known as peapods \cite{SmithNAT}, by a high
temperature treatment \cite{BandowCPL}.

The above mentioned chirality assignment method by Raman spectroscopy is at
first glance not automatically applicable for studying the properties of the
outer tubes and, in addition, seemingly suffers from several limitations
such as e.g. openings on the outer tubes are required. In addition, the
growth process of DWCNTs from fullerene peapods is not yet understood. From
computer simulation, it was demonstrated that C$_{60}$@SWCNT, peapod, based
DWCNTs are formed by Stone-Wales transformations from C$_{60}$ dimer
precursors formed at high temperature by cyclo-addition \cite%
{Tomanekprivcomm}\cite{SmalleyPRL}. The free rotation of C$_{60}$ molecules
is a prerequisite for the dimer formation as it enables the molecules to
have facing double bonds. It has been found experimentally that the
ellipsoidal shaped C$_{70}$ are present as both standing or lying peapod
configurations i.e. with the longer C$_{70}$ axis perpendicular or parallel
to the tube axis \cite{HiraharaPRB}\cite{KatauraAPA}.

In this Letter, we show that assigning the chiralities of the inner
nanotubes of DWCNT\ samples is a useful tool as the inner nanotube
distribution mimics that of the outer tubes. This allows the
characterization of diameter selective reactions or the measurement of the
nanotube abundance in the starting material. This is demonstrated on the
diameter selective healing of the SWCNT openings. Furthermore, the
comparison of C$_{60}$@SWCNT and C$_{70}$@SWCNT based DWCNTs evidences that
the distribution of inner tubes is very similar in the two kinds of samples.
A dramatic exception is observed for the $d$ $\approx 0.67$ nm inner
nanotubes for which the corresponding outer tubes are on the border between
lying and standing C$_{70}$ configurations. Such inner tubes are nearly
absent in the C$_{70}$@SWCNT based DWCNTs. The presence of very small inner
nanotubes, that can only be based on lying C$_{70}$, challenges the current
theorethical models of inner tube formation.


We prepared C$_{60}$,C$_{70}$@SWCNT based DWCNTs (60-DWCNT and 70-DWCNT,
respectively) from different SWCNT materials. Two arc-discharge prepared
commercial SWCNTs (SWCNT-N1 and N2 from Nanocarblab, Moscow, Russia \cite%
{nanocarblab}) and laser ablation prepared commercial (SWCNT-R from
Tubes@Rice, Houston, USA) and laboratory prepared SWCNT (SWCNT-L) were used.
The SWCNT-L samples are identical to those used previously \cite{PfeifferPRL}%
\cite{KrambergerPRB}. The SWCNT-N1, N2 materials were purified to 50 \% with
repeated high temperature air and acid washing treatments by the
manufacturer. SWCNT-R and SWCNT-L materials were purified following Ref. 
\cite{KatauraSM}. Peapod samples were prepeared by annealing SWCNT with C$%
_{60}$ in a quartz tube following Ref. \cite{KatauraSM} and were transformed
to DWCNT at high temperature following Ref. \cite{BandowCPL}. The diameter
distributions of the SWCNT materials were determined from Raman spectroscopy 
\cite{KuzmanyEPJB} giving $d_{\text{N}1}=$ 1.50 nm, $\sigma _{\text{N1}}$ =
0.10 nm, $d_{\text{N}2}=$ 1.45 nm, $\sigma _{\text{N}1}$ = 0.10 nm, $d_{%
\text{R}}=$ 1.35 nm, $\sigma _{\text{R}}$ = 0.09 nm, and $d_{\text{L}}=$
1.39 nm, $\sigma _{\text{L}}$ = 0.09 nm for the mean diameter and the
variance of the distribution for the different samples, respectively. We
have verified that the results described here are reproducible for all the
samples. Multi frequency Raman spectroscopy was performed on a Dilor xy
triple axis spectrometer in the 1.64-2.54 eV (755-488 nm) energy range and
in a Bruker FT-Raman spectrometer for the 1.16 eV (1064 nm) excitation at 90
K. The spectral resolution was 1-2 cm$^{-1}$ depending on the laser
wavelength. Raman shifts were calibrated against a series of spectral
calibration lamps.

\begin{figure}[tbp]
\includegraphics[width=0.8\hsize]{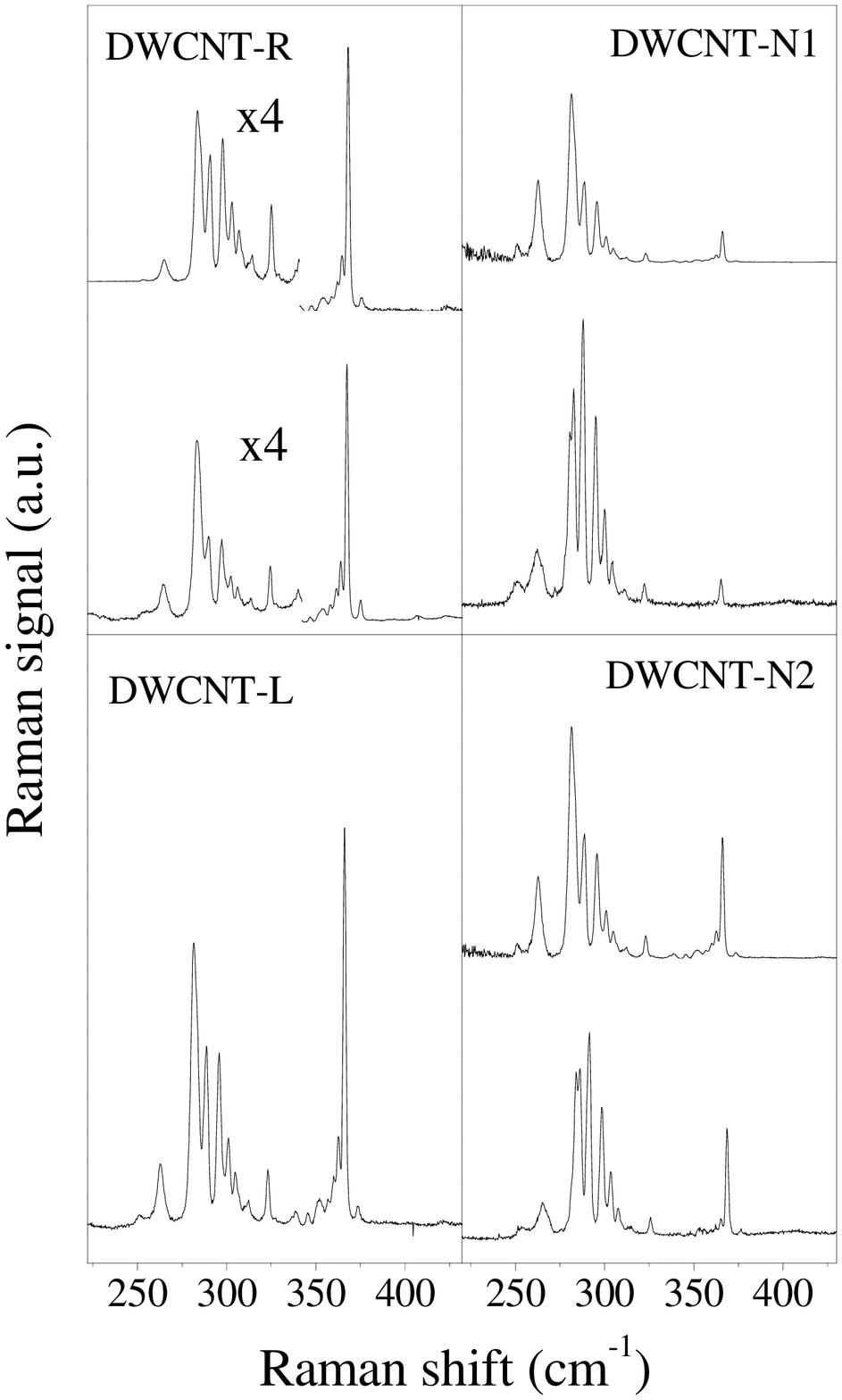} 
\caption{As measured Raman spectra of the inner nanotube RBMs for four
60-DWCNT samples (lower curves) at 647 nm laser
excitation. The upper spectra are smart scaled from the lower left spectrum
(DWCNT-L). The agreement between the as measured and smart scaled spectra
are obvious even though e.g. the N1, N2 tubes have clearly different
diameters.}
\end{figure}


In Fig. 1., we compare Raman spectra of the different 60-DWCNT materials for
a 647 nm excitation. The spectra are representative for the overall picture
observed for the measurements at other exciting laser energies and represent
the response from the radial breathing mode (RBM) of inner tubes only. The
splitting of the RBMs of geometrically allowed tubes is also observed in
Fig. 1. It was suggested to originate from one type of inner tubes being
encapsulated in outer tubes with slightly different tube diameters \cite%
{PfeifferPRL}. Interestingly, the RBMs of all the observable inner tubes,
including the split components, can be found at the same position in all
DWCNT samples within the $\pm $0.5 cm$^{-1}$ experimental precision of our
measurement for the whole laser energy range studied here. This agreement
between the different samples proves that vibrational modes of DWCNT samples
are robust against the starting material. This robustness goes beyond the
pure mode frequencies.

As the four samples have different diameter distributions, the overall Raman
patterns look different. However, scaling the patterns with the ratio of the
distribution functions allows to generate the overall pattern for all
systems, starting from e.g. DWCNT-L in the bottom-left corner of Fig.1. It
was assumed that the inner tube diameter distributions follow a Gaussian
function with a mean diameter 0.72 nm smaller than those of the outer tubes
based on a previous determination \cite{AbePRB} and with the same variance
as the outer tubes. We used the empirical constants from Ref. \cite%
{KrambergerPRB} for the RBM mode Raman shift versus inner tube diameter
expression. We observe a good agreement between the experimental and
simulated patterns for the DWCNT-R sample. A somewhat less accurate
agreement is observed for the DWCNT-N1, N2 samples, which may be related to
the different growth method: arc discharge, as compared to laser ablation
for the R and L samples. This agreement has important consequences for the
understanding of the inner tube properties. As a result of the
photoselective property of the Raman experiment, it proves that electronic
structure of the inner tubes is identical in the different starting SWCNT
materials. In addition, the diameter distribution of the inner tubes mimics
that of the outer ones, a fact that has been speculated previously \cite%
{PfeifferPRL}\cite{BandowCPL}\cite{AbePRB} and is established here. These
properties of the inner tubes make them excellent probes of the diameter
distribution of starting SWCNT materials, diameter selected SWCNTs and as
shown in the following, motivates the study of diameter selective properties
of the outer tubes.

\begin{figure}[tbp]
\includegraphics[width=0.8\hsize]{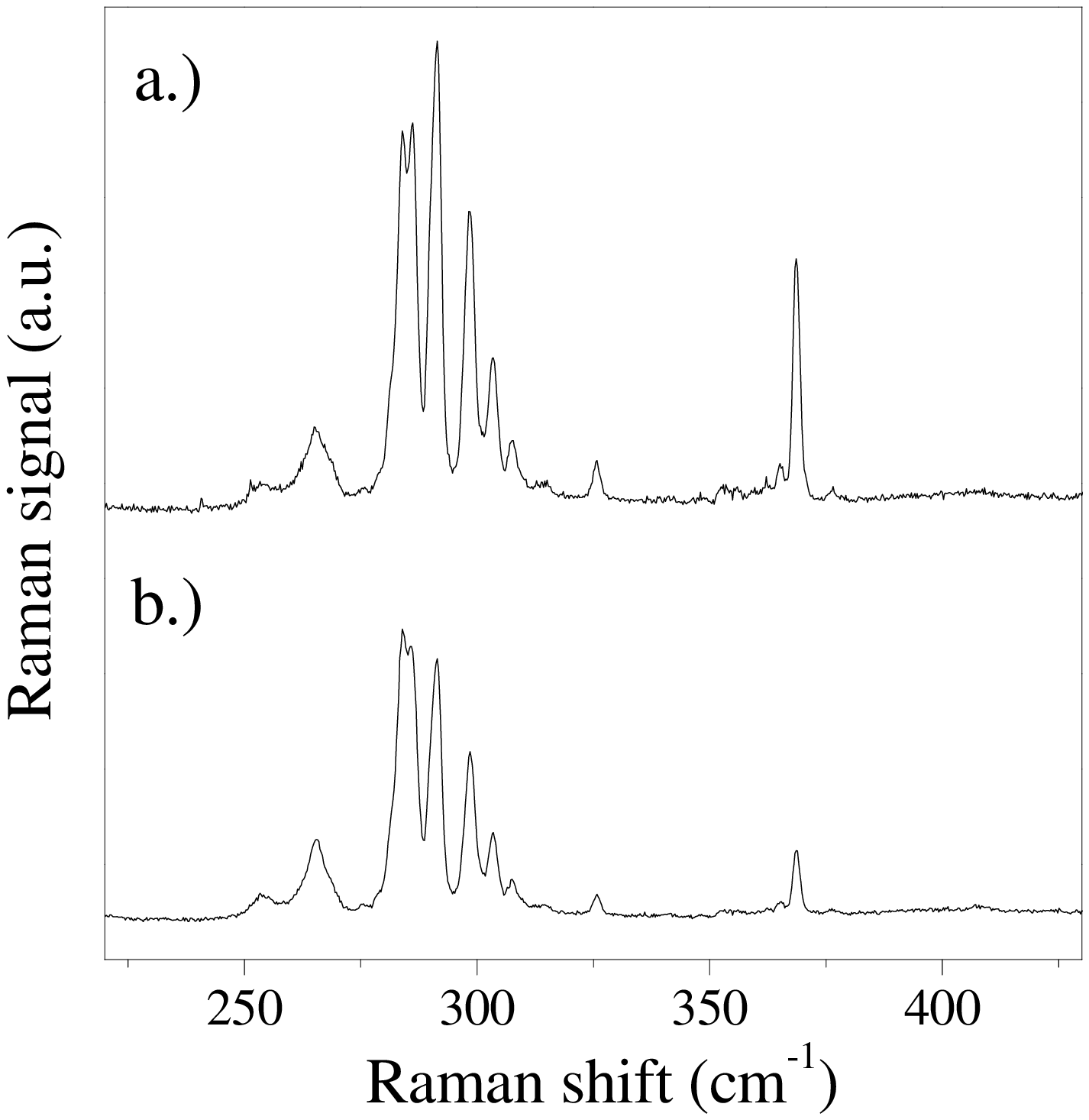} 
\caption{Raman spectra of a.) untreated DWCNT-N2, b.) 800 $^{o}$C
prior to filling treated DWCNT-N2 sample at 647 nm
excitation.}
\end{figure}

In Fig. 2., we show the Raman spectra of the 60-DWCNT-N2 heat treated at 800 
$^{\text{o}}$C in dynamic vacuum for 40 minutes prior to the C$_{60}$
encapsulation and DWCNT transformation as compared to an untreated sample.
The spectra are normalized by the amplitude of the corresponding outer
tubes. The weaker response at higher Raman shifted lines, i.e. inner tubes
with smaller diameters is apparent in Fig. 2b. This is related to the
annealing assisted closing of the outer tubes with small diameters, which
prevents the C$_{60}$ encapsulation. This effect is related to the higher
reactivity for healing of openings in small diameter SWCNTs, which provide
the host for the narrower inner tubes. Although, the overall healing effect
and the more rapid closing of smaller nanotubes has been long anticipated,
to our knowledge this is the first example when it is observed with an
individual tube sensitivity. The technique allows the accurate determination
of the healing speed of the different SWCNTs. A systematic study will be
published separately\cite{Hasiunpub}.

\begin{figure}[tbp]
\includegraphics[width=0.8\hsize]{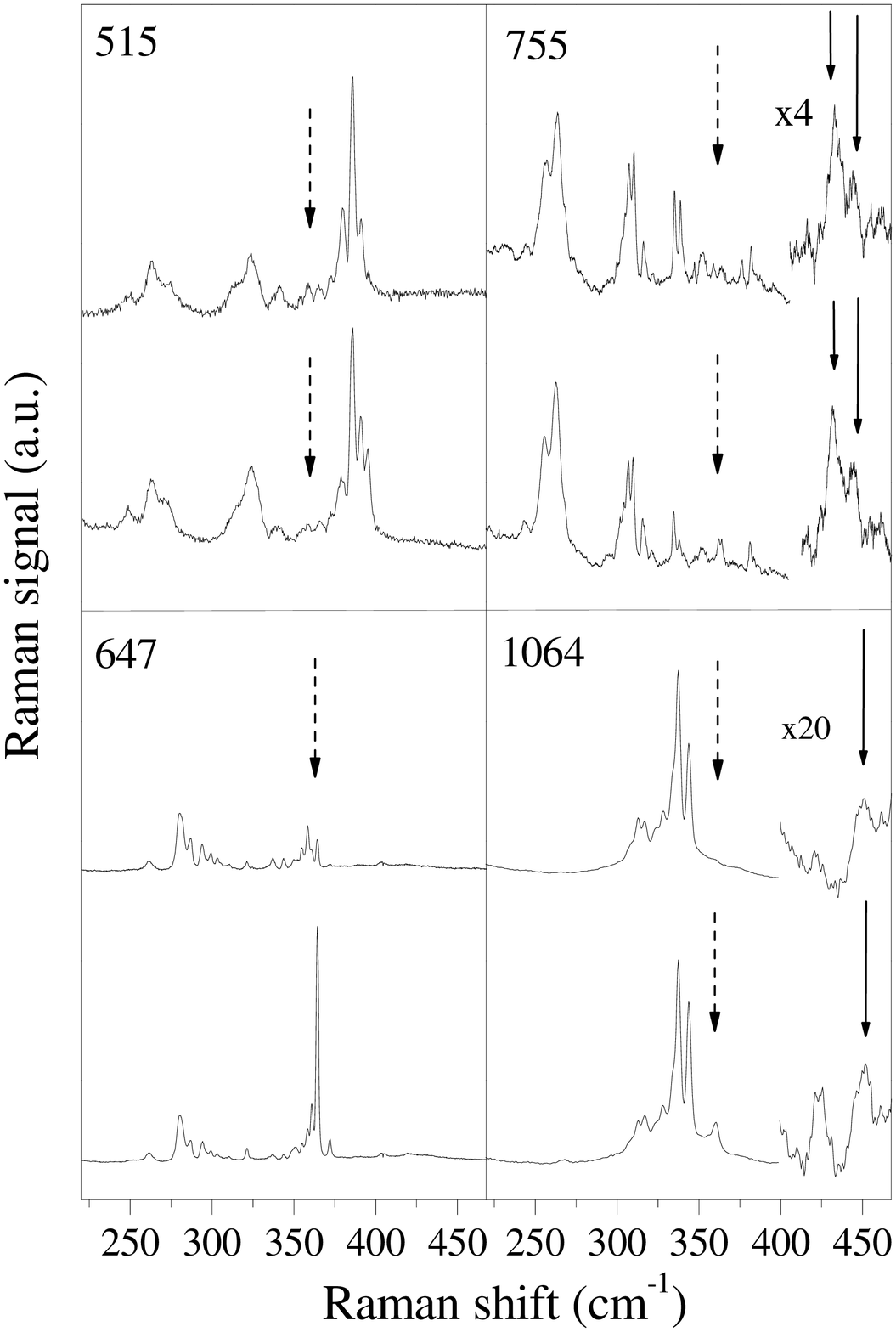} 
\caption{ Comparison of the Raman spectra of 60-DWCNT-R (lower spectra) and
70-DWCNT-R (upper spectra) for $\protect\lambda $=515, 647, 755, and 1064
nm, respectively. Solid arrows indicate the (7,5) and (5,3) nanotubes in the
755 nm spectra and the (4,4) nanotube on the 1064 spectra. Dashed arrow mark
the vicinity of the 362$\pm $3 cm$^{-1}$ spectral range.}
\end{figure}

Fig. 3. compares the Raman spectra of 60- and 70-DWCNT-R for some
representative laser energies. The RBMs of all the observable inner tubes,
thus including the split components, can be found at the same position for
all the 60-DWCNT and 70-DWCNT samples within our experimental precision.
This reinforces the previous finding that the inner tube formation is robust
against the starting SWCNT material or fullerene. The spectra shown in Fig.
3. are pair-wise normalized by the intensity of a selected inner tube from
the 300-340 cm$^{-1}$ spectral range. The photoselective property of the
Raman experiments prevents the use of the same normalizing inner tube for
all laser excitation energies. The observed similar pattern for 60-,
70-DWCNT proves that the electronic structure of the two kinds of DWCNT
materials is identical.

The observation of the very small inner tubes, with Raman shifts ranging up
to 430 cm$^{-1}$ for 70-DWCNT is important. It was previously found \cite%
{PfeifferPRL} that the smallest observable inner tubes for 60-SWCNT are the
(7,0), (5,3), (4,4) and (6,1) with diameters of 0.553, 0.553, 0.547, and
0.519 nm, respectively \cite{KrambergerPRB}. As indicated by solid arrows in
Fig 3., we clearly observe the (7,0), (5,3) and (4,4) for the 70-DWCNT-R
sample with intensities similar as in 60-DWCNT. The identification of the
(6,1) tube is less certain as it appears with very small intensity already
for the 60-DWCNT sample in the previous report \cite{PfeifferPRL}. Using the
experimentally determined 0.72 nm inner and outer tube diameter difference%
\cite{AbePRB}, the cut-off of the inner tube distribution at the (6,1) tube
for 60-DWCNT can be related to the smallest outer tube with $d_{\text{%
cut-off, C}_{\text{60}}}$ $\approx $ 1.239 nm where C$_{60}$ can enter. This
value is in reasonable agreement with theoretical estimates where $d_{\text{%
cut-off, C}_{\text{60}}}$ $\approx $ 1.2 nm was found for the smallest tube
diameter where C$_{60}$ peapod formation is energetically favored \cite%
{ZerbettoJPC}\cite{TomanekPRL}\cite{RochefortCM}\cite{OkadaPRB}. Similarly,
the energetics of the C$_{70}$ encapsulation was calculated and $d_{\text{%
critical}}$ $\approx 1.35$ nm was found for the SWCNT diameter which
separates the standing and lying configurations\cite{OkadaNJP}. Based on
this value, inner tubes with $d$ $\lesssim $ 0.63 nm can only be formed from
C$_{70}$ peapods in the lying configuration. The diameter of the above
mentioned and arrow-indicated tubes in Fig. 3. are all well below this
value. A small contamination of C$_{60}$ of the C$_{70}$ starting material
can be excluded to explain for the observed smallest diameter inner tubes as
it is unlikely to give rise to as many inner tubes as observed for the
60-DWCNT. Thus, it is evident that the smallest observed inner tubes for the
70-DWCNT are made nominally from lying C$_{70}$ peapod molecules.

The above result has important implications on the theoretical models of the
inner tube formation. It has been suggested that the route to inner tube
growth is the formation of cyclo-additonally bonded precursor C$_{60}$ dimers%
\cite{Tomanekprivcomm}\cite{SmalleyPRL}. Once the dimers are formed,
Stone-Wales transformations proceed till the completely formed inner tubes
are developed. However, the lying C$_{70}$ molecules are geometrically
hindered to form cyclo-additional dimers. Thus, the presence of inner tubes
from lying C$_{70}$ can not be explained by this model. Therefore, a
different process must be anticipated for the formation of very small
diameter inner tubes. If, however, an alternative mechanism is operational
for inner tube growth from 'lying' C$_{70}$, it is reasonable to assume that
the whole theory of inner tube formation requires revision. As an
alternative possibility for the formation of inner tubes, a complete decay
of the fullerenes into e.g. C$_{2}$ units may take place. This would be
consistent with the observation that the particular geometry of the given
fullerene does not play a role.

In what follows, an anomalous behavior observed for 70-DWCNTs with $d_{\text{%
outer}}\approx d_{\text{critical}}$ is discussed. The dashed arrows in Fig.
3. mark the vicinity of the 363$\pm 3$ cm$^{-1}$ Raman shifted RBMs which
were previously indentified to originate from the (5,5) metallic ($d$ = 0.68
nm) and the (8,1)\ semiconducting ($d$ = 0.671 nm) inner tubes along with
their split components \cite{PfeifferPRL}\cite{KrambergerPRB}. For a 647 nm
excitation an unusual and unexpected behavior is observed. Within the Raman
shift range indicated by the dashed arrows, some inner tube RBM components
are significantly weaker for the 70-DWCNT samples as compared to the
60-DWCNT samples. This is consistent with the results of the 755 and 1064 nm
excitations where weaker RBM modes are observed for the 70-DWCNT-R sample.
We observed the same anomalous behavior for the 70-DWCNT samples prepared
from the other SWCNT materials.

\begin{figure}[tbp]
\includegraphics[width=0.8\hsize]{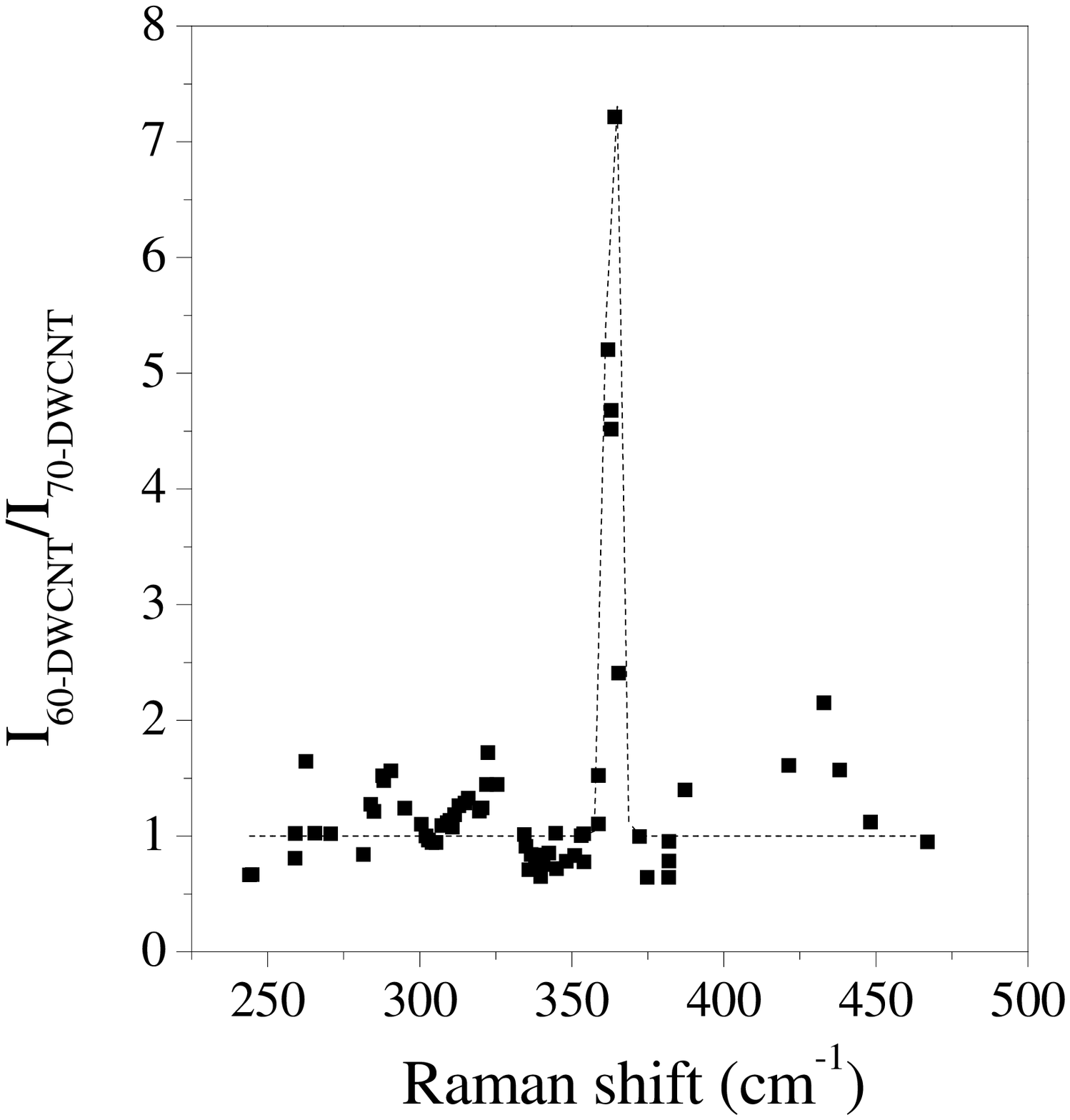} 
\caption{Normalized intensity ratios of the Raman RBM modes of inner tubes
in 60-, and 70-DWCNT-R materials at all laser lines studied. The data are
sliding averaged as explained in the text. The dashed curve is guide to the
eye.}
\end{figure}

To quantify this effect, we compared the intensity ratio of the RBM modes of
the inner tubes in 60-, and 70-DWCNTs for all the measured laser lines and
for all observed tubes. Voigtian lines were fitted to the observed spectra
with a Gaussian width determined by the spectrometer response to the
elastically scattered light (typically 1 cm$^{-1}$) and with a Lorentzian
line-width of the corresponding RBM mode. As discussed above, we chose a
particular tube in the 300-340 cm$^{-1}$ spectral range to normalize the
observed inner tube RBM\ intensities before dividing the so-obtained values
for the 60-, and 70-DWCNT by each other. Data points were collected together
for the studied 8 laser energies and RBM modes, sorted by the RBM frequency
and smoothed with a 3-point sliding averaging. This procedure reduces the
noise of the intensity ratio and makes a data point more reliable when the
same tube is observed at different laser energies. The result is summarized
in Fig. 4. The anomaly of the intensity ratio is clearly observed in the 363$%
\pm 3$ cm$^{-1}$ spectral range.

This spectral range corresponds to inner tubes with $d_{\text{inner}}\approx
0.67$ nm and $d_{\text{outer}}\approx $1.39 nm. As this latter value is
close to the critical diameter separating the lying and standing C$_{70}$
peapod configurations, it is tempting to associate the anomaly with the
competition between the two configurations. This, $d_{\text{critical}%
}\approx $ 1.39 nm, value refines the theoretical result of $d_{\text{%
critical}}\approx $ 1.35 nm. We discuss two possible origins of the missing
inner tubes: i) inability to form inner tubes from peapods filled with mixed
lying and standing C$_{70}$ configurations, ii) inability of C$_{70}$ to
enter into SWCNTs with $d_{\text{outer}}\approx d_{\text{critical}}$. It is
possible that SWCNTs with $d_{\text{outer}}\approx d_{\text{critical}}$ are
filled with mixed lying and standing C$_{70}$ molecules. The previously
proposed inner tube formation mechanism through dimer precursors followed by
Stone-Wales transformations may explain the absence of inner tubes formed
from such a mixture as the dimer formation might be very sensitive for the
mutual alignment of two adjacent C$_{70}$ molecules. However, we have shown
that inner tubes are also formed from lying C$_{70}$ molecules alone, which
disfavors the idea that the precursor dimer is indeed necessary for the
inner tube formation. Thus, it is speculated that the absence of inner tubes
when $d_{\text{outer}}\approx d_{\text{critical}}$ is rather caused by the
absence of C$_{70}$ molecules for peapods with this diameter.

A mechanism involving impurities or side-wall defects can explain our
observation: for the critical outer tube diameter separating the standing
from the lying configuration one can expect that for a perfect tube the
lying configuration is preferred. However, when side-wall defects or
impurties are present, a C$_{70}$ may change its configuration to a locally
preferred standing one that may immobilize it thus preventing other C$_{70}$
from entering into the tubes. Alternatively, a C$_{70}$ molecule entering at
the critical diameter may get trapped at a tube defect as e.g. a bend or
kink and thus prevents further filling. In addition, the competition between
elastic energy and rotation degrees of freedom may also give rise to a
blocking of C$_{70}$ in the tubes with $d_{\text{outer}}\approx d_{\text{%
critical}}$. The described diameter selective filling may provide a way to
mass-separate the unfilled peapod tubes from the filled ones. Interestingly,
the outer tubes which remain unfilled are close to the well studied (10,10)
tube.


In conclusion, DWCNT formation from peapods enables the study of diameter
selective phenomena in SWCNT materials. The method provides a new and
accurate tool for the characterization of directed nanotube growth, and the
effects of subsequent treatments or diameter selective separation. The
diameter selective closing of tube openings was observed for the first time.
Comparison of C$_{60}$,C$_{70}$ peapod based DWCNT proves that the inner
tube formation is a conservative process against the starting SWCNT or
fullerene material. The presence of very small inner nanotubes in 70-DWCNT
presents a challenge to the current theoretical models. The absence of
mid-diameter inner tubes in 70-DWCNT is explained by the absence of C$_{70}$
peapods for the corresponding nanotube diameter due to the borderline
between the lying and standing C$_{70}$ configurations.

This work was supported by the Austrian Science Funds (FWF) project Nr.
14893 and by the EU projects NANOTEMP BIN2-2001-00580 and PATONN Marie-Curie
MEIF-CT-2003-501099 grants and by the Zoltan Magyary Fellowship. This study
was partly supported by the Industrial Technology Research Grant Program in
'03 from the New Energy and Industrial Technology Development Organization
(NEDO) of Japan.

$^{*}$ Present address: Department of Applied \& Environmental Chemistry,
University of Szeged, Szeged, Hungary


\begin{thebibliography}{99}
\bibitem{IijimaNAT} S. Iijima, Nature (London) \textbf{354}, 56 (1991).

\bibitem{Dresselhaus} M. S. Dresselhaus, G. Dresselhaus, P. C. Ecklund: 
\textit{Science of Fullerenes and Carbon Nanotubes}, Academic Press, San
Diego 1996.

\bibitem{ChattopadhyayJACS} D. Chattopadhyay \textit{et al.}, J. Am. Chem.
Soc. \textbf{125}, 3370 (2003).

\bibitem{KrupkeSCI} R. Krupke \textit{et al.}, Science \textbf{301}, 344
(2003).

\bibitem{RinzlerNL} Z. H. Chen \textit{et al.}, Nano Lett. \textbf{3}, 1245
(2003).

\bibitem{StranoSCI} M. Zheng \textit{et al.}, Science \textbf{302}, 1545
(2003).

\bibitem{WeismanSCI} M. J. O'Connell \textit{et al.}, Science \textbf{297},
593 (2002).

\bibitem{PfeifferPRL} R. Pfeiffer \textit{et al.}, Phys. Rev. Lett. \textbf{%
90} 225501 (2003).

\bibitem{KrambergerPRB} Ch. Kramberger \textit{et al.}, Phys. Rev. B \textbf{%
68}, 235404 (2003).

\bibitem{SmithNAT} B. W. Smith, M. Monthioux, and D. E. Luzzi, Nature
(London) \textbf{396} 323 (1998).

\bibitem{BandowCPL} S. Bandow \textit{et al.}, Chem. Phys. Lett. \textbf{337}
48 (2001).

\bibitem{Tomanekprivcomm} D. Tom\'{a}nek, private communication.

\bibitem{SmalleyPRL} Y. Zhao, B. I. Yakobson, R. E. Smalley, Phys. Rev.
Lett. \textbf{88}, 185501 (2002).

\bibitem{HiraharaPRB} K. Hirahara \textit{et al.}, Phys. Rev. B. \textbf{64}%
, 115420 (2001).

\bibitem{KatauraAPA} H. Kataura \textit{et al.}, Appl. Phys. A \textbf{74},
349 (2002).

\bibitem{nanocarblab} http://www.nanocarblab.com

\bibitem{KatauraSM} H. Kataura \textit{et al.}, Synth. Met. \textbf{121},
1195 (2001).

\bibitem{KuzmanyEPJB} H. Kuzmany \textit{et al.}, Eur. Phys. J. B \textbf{22}%
, (2001) 307.

\bibitem{AbePRB} M. Abe \textit{et al.}, Phys. Rev. B \textbf{68}, (2003)
041405.

\bibitem{Hasiunpub} F. Hasi \textit{et al.}, unpublished.

\bibitem{ZerbettoJPC} M. Melle-Franco \textit{et al.}, J. Phys. Chem. B, 
\textbf{109}, 6986 (2003).

\bibitem{TomanekPRL} S. Berber, Y-K. Kwon, D. Tom\'{a}nek, Phys. Rev. Lett. 
\textbf{88}, 185502 (2002).

\bibitem{RochefortCM} A. Rochefort, cond-mat/0301310.

\bibitem{OkadaPRB} M. Otani, S. Okada, and A. Oshiyama, Phys. Rev. B \textbf{%
68}, 125424 (2003).

\bibitem{OkadaNJP} S. Okada, M. Otani, and A. Oshiyama, New J. Phys. \textbf{%
5}, 122 (2003).
\end{thebibliography}
\end{document}